\documentclass[a4paper, 10pt, conference]{ifacconf}      

\usepackage{amsmath}
\usepackage{amsfonts}
\usepackage{amssymb}
\usepackage{exscale}
\usepackage{graphicx}
\usepackage{color}

\usepackage{algorithmic}
\usepackage{graphicx}
\usepackage{gensymb}
\usepackage{multirow}
\usepackage{array}
\usepackage{arydshln} 
\usepackage{makecell}

\usepackage[utf8]{inputenc}
\usepackage{soul}
\usepackage{censor}
\usepackage{url}
\usepackage{epsfig}
\usepackage{lscape}
\usepackage{todonotes}								
\usepackage{subeqnarray}
\usepackage{calc,array}
\usepackage{import}
\usepackage{upgreek}
\usepackage{bm}

\usepackage{natbib}        

\usepackage{fancybox}

\usepackage{booktabs}

\usepackage{epstopdf}

\usepackage{array}

\newcommand{\D}{\mathcal{D}}

\usepackage{textcomp}
\usepackage{mathdots}
\usepackage{siunitx}
\usepackage{tikz}
\usetikzlibrary{arrows, positioning, calc, fit}

\usepackage{environ}

\makeatletter
\newsavebox{\measure@tikzpicture}
\NewEnviron{scaletikzpicturetowidth}[1]{%
	\def\tikz@width{#1}%
	\def\tikzscale{1}\begin{lrbox}{\measure@tikzpicture}%
		\BODY
	\end{lrbox}%
	\pgfmathparse{#1/\wd\measure@tikzpicture}%
	\edef\tikzscale{\pgfmathresult}%
	\BODY
}
\makeatother

\makeatletter
\def\endfigure{\end@float}
\def\endtable{\end@float}
\makeatother

\let\ifacconfcaptionwidth\captionwidth
\usepackage[caption=false]{subfig}
\let\captionwidth\ifacconfcaptionwidth

\renewcommand{\j}{\mathrm{j}}
\newcommand{\hinf}{{$\mathcal{H}_\infty$} }

\begin{document}
\begin{frontmatter}
\title{Comparison of Fractional-Order and Integer-Order $\bm{\mathcal{H}_\infty}$ Control\\of a Non-Collocated Two-Mass Oscillator}

\author[First]{Benjamin Voß}
\author[Third]{Michael Ruderman}
\author[First]{Christoph Weise}
\author[First]{Johann Reger\thanksref{Fourth}} 

\address[First]{Control Engineering Group, Technische
 Universit\"at Ilmenau, P.O. Box 10 05 65, D-98684, Ilmenau, Germany}%
\address[Third] {Department of Engineering Sciences, University of Agder,
  P.O. Box 422, 4604 Kristiansand, Norway, {\tt\small michael.ruderman@uia.no}}

\thanks[Fourth]{Corresponding author: {\tt\small johann.reger@tu-ilmenau.de}\\\copyright ~2022 the authors. This work has been accepted to IFAC for publication under a Creative Commons Licence CC-BY-NC-ND.}%

\begin{abstract}
We consider the robust control of a two-mass oscillator with a dominant input delay. Our aim is to compare a fractional-order tuning approach including the partial compensation of non-minimum phase zeros with a classical $\mathcal{H}_\infty$ loop-shaping design, since both these designs lead to a relatively high controller order.
First of all a detailed physical model is derived and validated using measurement data. Based on the linearized model both controllers are designed to be comparable, i.e. they show a similar crossover frequency in the open loop and the final controller order is reduced to the same range for both designs. 
The major differences between both are the different methods how the feed-forward action is included. The loop-shaping approach with fractional-order elements relies on the plant inverse using a flat output, whereas the $\mathcal{H}_\infty$ design incorporates a two-degree of freedom control, i.e. the reference signal is included into the known inputs of the generalized plant. Each controller is tested in simulation and experiment. As both open-loops are nearly identical in the frequency range of interest, the results from an input disturbance experiment show no major difference. The different design approaches of the feed-forward path are clearly visible in the tracking experiment.

\end{abstract}

\end{frontmatter}

\section{Introduction}

With many numerical tools available \citep{TepljPB11,oustaloup_crone_2000,Valerio04-NIinteger}, the use of fractional-order~(FO) operators in control loops has gained popularity. Especially the extension of the classical PID-controller \citep{Podlubny99-PID} by FO terms has been investigated intensively \citep{MonjeCVXF10}, as the FO integral part leads to a reduced phase loss compared to an integer-order~(IO) integrator. In addition to that the scalable slopes in the amplitude response offer a high degree of freedom in the frequency domain with a minimal number of parameters to tune. 

A drawback of these FO design concepts yet is the implementation. As FO operators are non-local, a lot of physical memory is necessary, i.e. a higher-order IO approximation, to apply these control strategies in real-time. In most contributions an FO controller approach is only compared to standard PID controllers, despite the fact that the controller orders are hardly comparable, e.g. a reasonable approximation of an FO term requires five states, whereas a realizable PID-controller only needs two. As all of various approximations schemes lead to controllers of a high order, it is natural to compare the performance of these FO approaches with design approaches leading to a similar number of states, e.g. the $\mathcal{H}_\infty$ framework. A first attempt for such a comparison can be found in \citep{seyedtabaii_modified_2019}. 

In this paper we consider the control of a two-mass oscillator with a nominal input delay. We design two higher-order controllers and compare the resulting tracking- and disturbance performance. The first controller approach is based on an open-loop PI controller, which is extended by a partial FO compensator and a notch filter. In this case the feed-forward control is designed based on the model inverse. The approximation of the FO part is chosen such that the overall controller order does not exceed order 7. For comparison, a two-degree of freedom~(2DOF) $\mathcal{H}_\infty$ controller is applied. With similar open loop-characteristics both controllers show the expected disturbance rejection behavior. The major difference can be seen in the tracking experiment. The model inverse leads to a faster response, but the effect of an unknown input delay is more severe as the overshooting changes significantly.

The structure of this contribution is as follows. In Section~\ref{sec:sys_dynamics} a detailed description of the experimental setup is given and a model of the two-mass oscillator is derived. 
As a focus of this paper is to investigate the partial cancellation of non-minimum phase zeros, a communication delay is included and approximated by a Padé-term. The controller design is explained in Section~\ref{sec:controllerdesign}. We compare an open-loop design including an FO partial compensator with an IO $\mathcal{H}_\infty$ controller. Both controllers show a nearly identical open loop within the frequency range of interest. However, the different approaches to include the feed-forward action lead to significant differences regarding the robustness and tracking performance. The simulation and experimental results are given in Section~\ref{sec:results}. Besides the tracking test with different time delays, both controllers are also subjected to an artificially injected band-limited white noise. Section~\ref{sec:conclusions} concludes the paper.

\section{System Dynamics}
\label{sec:sys_dynamics}

\begin{figure}[ht]
	\centering
	\includegraphics[width=.7\linewidth]{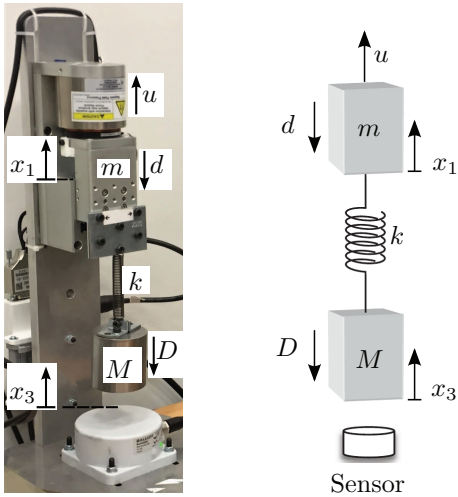}
	\caption{Experimental setup of the two-mass oscillator \citep{Ruderman2021,Ruderman2022}.}
	\label{fig:schematics-experiment}
\end{figure}

The considered plant is a two-mass oscillator with non-collocated actuator (MGV~52~\citep{AkribisVCM}) and measurement (BAW003K~\citep{Balluff2021}). An illustration of the system in the laboratory setting is given in Fig.~\ref{fig:schematics-experiment}. The control signal ${u(t)\in [0\,\text{V},\,10\,\text{V}]}$ leads to a force at the active mass $m$ that is connected to the load (passive mass $M$) by a spring with stiffness constant $k$. The displacement of the active mass is bounded within ${0\,\text{mm} \leq x_1(t) \leq 20\,\text{mm}}$. Both the active and the passive masses are subject to gravity ${d=-mg}$ and ${D=-Mg}$, respectively, where $g$ is the gravity constant. Only the position of the hanging load $\tilde{x}_3(t)$ is available for control by contact-less measurement and is used in relative coordinates $x_3(t) =  \tilde{x}_3(t)-\tilde{x}_{3,0}$, where $\tilde{x}_{3,0}$ is the steady-state measurement for $u=0$. 

Since the actuator inductance is relatively small~\citep{AkribisVCM}, the current dynamics can be neglected. This results in a model similar to \citep{Ruderman2021} with state-space representation
\vspace{-1.2ex}
\begin{equation}\label{eq:systemdynamics-ssr}
\dot{x} = A{x} + B {u} + E\begin{bmatrix}d\\D \end{bmatrix}  ,\quad y = C {x} ,\quad x(t_0) = {x}_0 
\end{equation}
with the state vector ${x = \begin{bmatrix} x_1 & {x}_2& x_3 & x_4\end{bmatrix}^\top}$ where ${x_2 = \dot{x}_1}$ and ${x_4 = \dot{x}_3}$, initial conditions ${x}_0$, the input and output vectors ${B = \begin{bmatrix}0&\frac{\Psi}{Rm}&0&0\end{bmatrix}^\top}$ and ${C = \begin{bmatrix}0&0&1&0\end{bmatrix}}$, respectively, as well as the matrices
\begin{equation*}
A = \begin{bmatrix}
0& 1& 0& 0\\
    \frac{-k}{m}&\frac{-1}{m}\left(\eta+\zeta\right)& \frac{k}{m}& \frac{\zeta}{m}\\
    0 &0 &0 &1\\
    \frac{k}{M}&\frac{\zeta}{M}&  \frac{-k}{M} &\frac{-\zeta}{M}
\end{bmatrix}
,\, E = \begin{bmatrix}
 0 &0 \\
 \frac{1}{m}&0\\
0&0\\
0&\frac{1}{M}
\end{bmatrix}.\vspace{12pt}
\end{equation*}
The actuator properties $\Psi$ and $R$ are the back electromotive force (EMF) constant and resistance, $\eta$ and~$\zeta$ characterize the damping of the active mass and connecting spring, respectively, while $\zeta \ll \eta$. 
As the gravitational forces are assumed to be constant and known, they can be compensated by feed-forward action $u_d = -(d+D)\frac{R}{\Psi} = {(m+M)\frac{gR}{\Psi}}$, if we choose the coordinate system to be located at the equilibrium point with respect to $D$. Thus, they are not part of further discussions.
\begin{table}[ht]
	\centering
	\caption{Nominal values of model parameters.}
	\label{tab:systemdynamics-parameters}
	\begin{tabular}{r|lll}
		\textbf{Parameter}	&	\textbf{Value}	&	\textbf{Unit}	&	\textbf{Meaning}\\ \hline
		$k$						&	200				&	N/m				&	Spring constant\\
		$m$						&	0.6				&	kg					&	Active mass\\
		$M$						&	0.75				&	kg					&	Passive mass (load)\\
		$\eta$				&	200				&	kg/s				&	Visc. damping of active mass\\
		$\zeta$					&	0.02				& 	kg/s				&	Visc. damping of conn. spring\\
		$\Psi$						&	17.16				&  Vs/m				&	Actuator EMF constant\\
		$R$						&	5.23				&  V/A				& Actuator resistance\\	
		$g$						&  9.81				&  m/s$^2$		& Gravitational acceleration
	\end{tabular}
\end{table}
By utilizing the state-space realization~\eqref{eq:systemdynamics-ssr} and the nominal values of the parameters, given in Table~\ref{tab:systemdynamics-parameters}, the transfer function is
\begin{equation}\label{eq:systemdynamics-Gyu}
\begin{aligned}
G_\mathrm{yu}(s) 	&= C\left(sI-A\right)^{-1}B\\
				&= \frac{0.14583\, (s+10^4)}{s\, (s+332.4)\, (s^2 + 1.027s + 267.4)},
\end{aligned}
\end{equation}
where $I$ denotes the identity matrix of appropriate dimensions. Apart from a stable fast pole ($p_1 = -332.4$) and zero, the dominant dynamics are described by the free integrator ($p_2 = 0$) as well as the low-damped complex pole-pair at $p_{3,4} = -0.514 \pm \j\,16.346$.

\begin{figure}[ht]
	\centering
	\includegraphics[width=\linewidth]{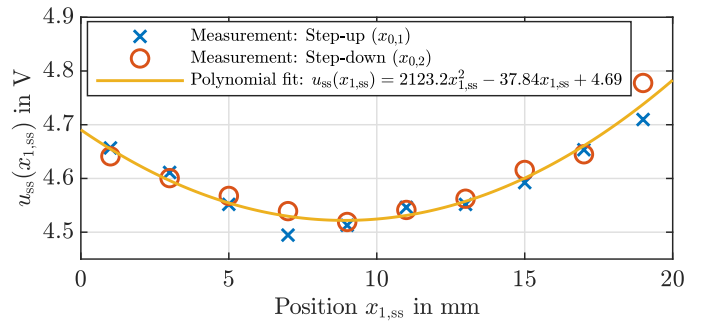}
	\caption{Experimental identification of the steady-state input voltage to compensate the gravitational forces.}
	\label{fig:systemdynamics-uss}
\end{figure}

A symmetrical open-loop excitation has revealed a good fit of the eigenfrequency~$\omega_0 = 16.346\,\mathrm{\frac{rad}{s}}$, however major differences in the transient peaking and damping. 
Furthermore, the actuator property of a stroke-dependent force can be observed in terms of a state-dependent input gain~$k_u(x_1)$.
Closed-loop experiments have been conducted in order to identify~$k_u(x_1)$. As the active and the passive mass are connected with a spring (cf. Fig.~\ref{fig:schematics-experiment}), their relative displacement equals in steady-state: $x_{1,\mathrm{ss}}=x_{3,\mathrm{ss}}$. By utilizing a controller with integral action, steady-state tracking accuracy for a reference step can be achieved. Thus, given the steady-state reference value $r_\mathrm{ss}$ for the position of the passive mass, the necessary input signal can be linked to the active mass' position: $u_\mathrm{ss}=u_\mathrm{ss}(x_\mathrm{1,ss})$. For this purpose, a grid of reference positions is targeted from two opposite initial conditions $ {x}_{0,1} =   \left[0~0~0~0\right]^\top$ and ${ x_{0,2} =   \left[0.02~0~0.02~0\right]^\top}$.
The experimental results are summarized in Fig.~\ref{fig:systemdynamics-uss}.
Furthermore, the least-squares optimal, second order polynomial fit $u_\mathrm{ss}(x_\mathrm{1,ss})$
is shown. This allows calculating the input gain $k_u(x_1)$ by utilizing the equilibrium of forces in steady-state, i.e. the gravitational forces on the one hand and the actuator force on the other hand, leading to
\begin{equation}
k_{u}(x_1) = (m+M)gR (\Psi  u_\mathrm{ss}(x_1))^{-1}.\label{eq:systemdynamics-ku}
\end{equation}
Note that we cannot use the inverse of this nonlinearity because the position of the active mass is not measured. Best results are achieved by utilizing the mean of the inverse input gain $\mu\left(k_u^{-1}\right)$ as static pre-compensation. Still, $k_{u}(x_1)$ is utilized to improve the model for later simulation studies. 

In order to simulate a possible essential communication delay, a dominant input delay $\tau$ is added to the plant model. This results in the transfer function
\begin{equation}\label{eq:systemdynamics-G}
G(s) 	= G_\mathrm{yu}(s) ~ e^{-\tau s}\equiv G_\mathrm{yu}(s) G_\mathrm{d}(s) 
\end{equation}
with the nominal value for the delay $\tau_\mathrm{n} = \frac{2\pi}{\omega_0} = 0.3844\,\text{s}$.


\section{Controller Design}
\label{sec:controllerdesign}
In this section, two different synthesis methods are utilized to derive controllers for the plant model $G(s)$, provided in~\eqref{eq:systemdynamics-G}, including the dominant communication delay. On the one hand, \hinf controller synthesis methods are used to design a reference controller. On the other hand, FO loop shaping with classical feed-forward is utilized. A minimum phase margin of $30^\circ$ is to be achieved to allow for delay uncertainties. For a fair comparison, both controllers are designed to have a similar order~$n$, open-loop crossover frequency~$\omega_\mathrm{c}$ and gain for low frequencies. For good tracking performance, 2DOF control structures are used.

\subsection{Two Degree of Freedom $\mathcal{H}_\infty$ Control}
To be able to use the classical \hinf synthesis methods~\citep{Zhou1996}, we introduce
the second-order all-pass Padé-term $\tilde{G}_\mathrm{d,2} \approx G_\mathrm{d}$ from \cite{Pade1892} leading to
\begin{equation}
	\label{eq:controllerdesign-plant}
	\tilde{G}(s) = G_\mathrm{yu}(s)~ \tilde{G}_\mathrm{d},
\end{equation}
where the time delay for the controller design $\tau_\mathrm{n}$ coincides with the nominal delay of the plant. For better tracking performance a 2DOF control structure is utilized, see Fig.~\ref{fig:schematics-2dof-control-loop}, mainly inspired by \citep{Gu2005}.
\tikzstyle{block} = [draw, fill=white, rectangle, minimum height=2.5em, minimum width=3em, anchor=center]
\tikzstyle{sum} = [draw, fill=white, circle, minimum height=0.6em, minimum width=0.6em, anchor=center, inner sep=0pt]
\begin{figure}[ht]
\centering
	\begin{scaletikzpicturetowidth}{\linewidth}
		\begin{tikzpicture}[scale=\tikzscale]
		\node[coordinate](input) at (0,0) {};
		\node[block] (controller) at (2,0) {$C_\infty(s)$};
		\node[block] (system) at (6.8,0) {$G_\mathrm{yu}(s)$};
		\node[block] (timedelay) at (4.5,0) {$e^{-\tau s}$};
		\node[above] (G) at (7.3,.6) {$G(s)$};
		\node[draw,dashed,rectangle, fit=(timedelay) (system)] (extsys) {};
		\node[coordinate](output) at (9,0) {};
		\node[coordinate](disturbance1) at (5.6,.75) {};
		\node[coordinate](disturbance2) at (5.5,-1) {};
		\coordinate (aux1) at ($(controller.north west)!.3333!(controller.south west)$);
		\coordinate (aux2) at ($(controller.north west)!.6666!(controller.south west)$);
		\node[coordinate](aux3) at (1,0) {};
		\node[sum, fill=black, minimum size=0.4em] (dot1) at (8.3,0) {};
		\node[sum, fill=white, minimum size=0.4em] (sum1) at (5.6,0) {};
		\draw[thick,-latex] (disturbance1) node[above]{$d_u(t)$} --  (sum1);
		
		\draw[thick,-latex] (input|-aux1) -- node[above]{  $r(t)$} (aux1);
		\draw[thick,-latex] (controller) -- node[above]{  $u(t)$}(timedelay) ;
		\draw[thick,-latex](timedelay) -- (sum1);
		\draw[thick,-latex](sum1)-- (system);
		
		\draw[thick,-latex] (system)  -- node[above]{ $y(t)$} (output);
		\draw[thick,-latex] (dot1) -- (8.3,-1) -- (1,-1) -- (aux3|-aux2) -- (aux2);
		\end{tikzpicture}
	\end{scaletikzpicturetowidth}
	\caption{Two degree of freedom control structure.}
	\label{fig:schematics-2dof-control-loop}
\end{figure}
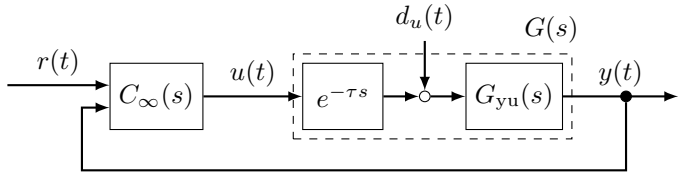

Furthermore, the disturbance~attenuation for input disturbances is of special interest.
For this purpose, the external input is extended to $w=\begin{bmatrix}r&d_u\end{bmatrix}^\top$.  The performance output 
\begin{equation}
z = \begin{bmatrix}z_1&z_2\end{bmatrix}^\top = \begin{bmatrix} W_1(y-r) & W_2 u\end{bmatrix}^\top
\end{equation}
takes the effect of the reference $r$ and the disturbance $d_u$ on the control error $e = (r-y)$ and the controller output~$u$ into account.  Leading to the generalized plant
\begin{equation}
	P = \left[\begin{array}{cc:c}
				-W_1 & W_1 \tilde G & W_1 \tilde G\\
				0 & W_2 & W_2\\\hdashline 
				1 & 0 & 0\\ 
				0 & \tilde G & \tilde G	
			\end{array}\right] ~ \text{with} ~			
	\begin{bmatrix} z_1\\ z_2\\ r\\y\end{bmatrix} = P(s) \begin{bmatrix} r\\ d_u\\ u\end{bmatrix},
\end{equation}
where the arguments are omitted for clarity reasons. 
As proposed by~\cite{Skogestad2001}, the weight functions are chosen as 
\begin{equation}
	W_1(s) = \frac{0.01(s+15)}{s+1.5\cdot10^{-3}}~\text{ and }~ W_2(s) = \frac{0.95 (s+3)}{s+2850}.
\end{equation}
The low-pass filter $W_1$ with high gain for low frequencies leads to an (almost) integrating behavior with a pole at~$-\epsilon$, whereas the high-pass filter $W_2$ limits the tracking performance but ensures noise reduction by limiting the bandwidth of $u$. 
\begin{figure*}[!t]
\normalsize
\begin{subequations}\label{eq:controllerdesign-Cinf}
	\begin{align}
		C_{\infty,1} (s) &= \frac{-0.030643 (s-538.6) (s+31.93) (s+8.538) (s^2 - 32.18s + 1.846\cdot 10^4) (s^2 - 314s + 1.923\cdot 10^5)}{s (s+507.2) (s+72.68) (s^2 + 9.086s + 63.62) (s^2 + 307.9s + 1.36\cdot 10^5)}\\
		C_{\infty,2}  (s) &= \frac{ -3.3043 (s-3.893\cdot 10^4) (s-1534) (s+62.69) (s+7.822) (s+0.6413) (s^2 + 0.5762s + 257.6)}{s (s+507.2) (s+72.68) (s^2 + 9.086s + 63.62) (s^2 + 307.9s + 1.36\cdot 10^5)}
	\end{align}
\end{subequations}
\begin{equation}\label{eq:controllerdesign-Cfo}
	C_\mathrm{FO} (s) = \frac{120.2 (s+14.09) (s+6.092) (s+5.292) (s+2) (s+0.6667) (s^2 + 1.027s + 267.4)}{s (s+6) (s+5.484) (s+5.203) (s+8.015) (s+9) (s+33.32)}
\end{equation}
\hrulefill
\vspace*{4pt}
\end{figure*}

\begin{figure}[ht]
	\centering
	\includegraphics[width=\linewidth]{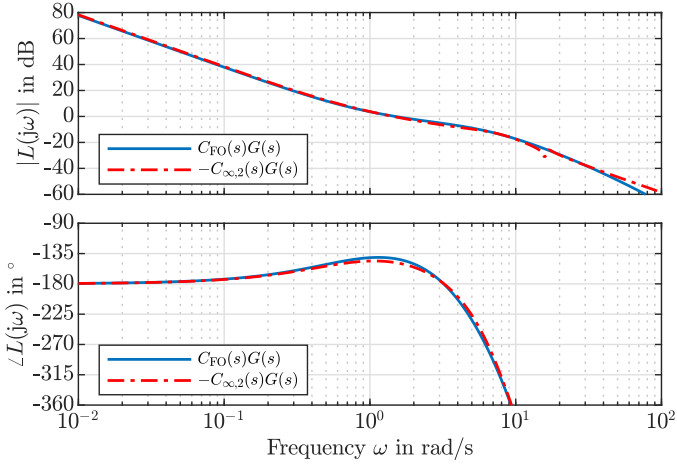}
	\caption{Bode plot of the open loop transfer functions.}
	\label{fig:controllerdesign-bode}
\end{figure}

Using the MATLAB Robust Control Toolbox function \texttt{hinfsyn}, a sub-optimal stabilizing 2DOF controller $\hat{C}_\infty$ is found determining
\begin{equation}
\left\| P_{11}+ P_{12}\hat{C}_\infty(s)\left(I-P_{22}\hat{C}_\infty(s) \right)^{-1} P_{21} \right\|_\infty < 1
\end{equation} 
for the transfer function from the external input $w$ to the performance output $z$. In order to achieve zero steady state error, an integrator is introduced
\begin{equation}
\hat{C}_\infty(s) = \frac{1}{\epsilon + s}\tilde{C}_\infty \approx \frac{1}{s}\tilde{C}_\infty \equiv C^\prime_\infty(s)
\end{equation}
where $\epsilon < 2\cdot 10^{-3}$. Finally, the controller order is reduced via balanced truncation (based on~\citep{Varga1991}) resulting in a controller order of $n_\infty = 7$ and the controller given by $		C_\infty = \left[C_{\infty,1}~C_{\infty,2} \right]$, see \eqref{eq:controllerdesign-Cinf} next page,
where $C_{\infty,1}$ and $C_{\infty,2}$ correspond to the reference $r$ and the output~$y$, respectively. For later comparison, a Bode plot of the open loop $L_\infty= -C_{\infty,2}G$ for $r=0$ is depicted in Fig.~\ref{fig:controllerdesign-bode}, showing a crossover frequency of $\omega_{\mathrm{c},\infty} = 1.44\,\mathrm{\frac{rad}{s}}$, a phase margin of $\Phi_\mathrm{{r},\infty}= 32.09^\circ$ and a gain margin of $A_\mathrm{{r},\infty}= 2.18$.

\begin{figure}[ht]
	\centering
	\includegraphics[width=\linewidth]{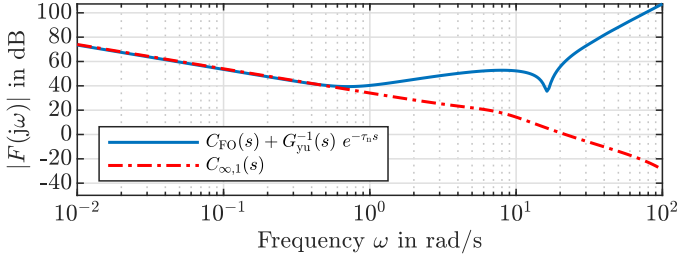}
	\caption{Magnitude plot of the feedforward paths.}
	\label{fig:controllerdesign-bode-prefilter}
\end{figure}

The magnitude plot of feedforward path $F_\infty = C_{\infty,1}$ from reference $r$ to control signal $u$ is shown in Fig.~\ref{fig:controllerdesign-bode-prefilter}. High frequencies of the reference signal are not amplified, whereas it shows integrating behavior for low frequencies. 

\subsection{FO Loop-Shaping with Classical Feedforward}
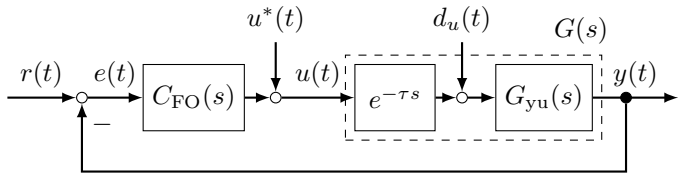
\begin{figure}[ht]
	\centering
	\begin{scaletikzpicturetowidth}{\linewidth}
		\begin{tikzpicture}[scale=\tikzscale]
		\node[coordinate](input) at (0,0) {};
		\node[sum, fill=white, minimum size=0.4em] (sum1) at (1,0) {};
		\node[block] (controller) at (2.5,0) {$C_\mathrm{FO}(s)$}; 
		\node[block] (system) at (7.2,0) {$G_\mathrm{yu}(s)$};
		\node[block] (timedelay) at (5.2,0) {$e^{-\tau s}$};
		\node[above] (G) at (7.7,.6) {$G(s)$};
		\node[draw,dashed,rectangle, fit=(timedelay) (system)] (extsys) {};
		\node[coordinate](output) at (9,0) {};
		\node[coordinate](feedforward) at (3.6,.75) {};
		\node[coordinate](disturbance1) at (6.1,.75) {};
		\node[coordinate](disturbance2) at (5.5,-1) {};
		\node[sum, fill=black, minimum size=0.4em] (dot1) at (8.3,0) {};
		\node[sum, fill=white, minimum size=0.4em] (sum2) at (6.1,0) {};
		\node[sum, fill=white, minimum size=0.4em] (sum3) at (3.6,0) {};
	
		\draw[thick,-latex] (disturbance1) node[above]{$d_u(t)$} --  (sum2);
		\draw[thick,-latex] (feedforward) node[above]{$u^*(t)$} --  (sum3);
		\draw[thick,-latex] (input)  -- node[above]{$r(t)$} (sum1);
		\draw[thick,-latex] (sum1)  -- node[above]{$e(t)$} (controller);
		\draw[thick,-latex] (controller) -- (sum3);
		\draw[thick,-latex] (sum3) -- node[above]{  $u(t)$}(timedelay);
		\draw[thick,-latex] (timedelay) -- (sum2);		
		\draw[thick,-latex] (sum2) -- (system);
		\draw[thick,-latex] (system)  -- node[above]{$y(t)$} (output);
		\draw[thick,-latex] (dot1)  |- (6,-1) -|  node[pos=0.85,right]{$-$} (sum1);
		\end{tikzpicture}
	\end{scaletikzpicturetowidth}
	\caption{Control structure with feedforward action~$u^*(t)$.}
	\label{fig:schematics-standart-control-loop}
\end{figure}

The classical PI-Lead controller design is combined with partial cancellation of the dominant non-minimum phase zero of $\tilde{G}$ and the integer-order~(IO) compensation of the conjugate complex pole-pair at ${\omega_0 = 16.346\,\mathrm{\frac{rad}{s}}}$. 
Figure~\ref{fig:schematics-standart-control-loop} shows the 2DOF control structure.

Using a first-order Padé-approximation $\tilde{G}_\mathrm{d,1} \approx G_\mathrm{d}$ from \cite{Pade1892} for time delay~$\tau_\mathrm{n} = 0.3844\,\mathrm{s}$ leads to a non-minimum phase zero in $\tilde{G}$ close to the desired crossover~frequency~$\omega_\mathrm{c}$ (see \eqref{eq:controllerdesign-plant} and Fig.~\ref{fig:controllerdesign-bode}). Following the reasoning of~\citep{Voss2022}, we partially compensate the non-minimum phase zero $z =2/\tau_\mathrm{n}$ with $\tilde Q_{2/\tau_\mathrm{n},2}$ of order ${\nu = \alpha^{-1} = 2}$, leading to a minimum phase lag and gaining some magnitude slope as well.

Designing a PI control for $\tilde Q_{2/\tau_\mathrm{n},2}\tilde{G}$ in order to reach a crossover frequency similar to $\omega_{\mathrm{c},\infty}$ reveals two difficulties. First, the damping of the magnitude peak at $\omega_0$ limits the applicable proportional gain significantly. Knowing that the model damping is over-estimated, this poses a major issue. However, since the ei\-gen\-frequency fits the model, we can use the non-proper integer-order filter 
\begin{equation}
F_\mathrm{pp}(s) = \frac{ 0.034\left(s^2 + 1.027s + 267.4\right)}{s+9}.
\end{equation}
Still, the combination of all elements yields a proper overall controller $C_\mathrm{FO}$, see~\eqref{eq:controllerdesign-Cfo}. The frequency of the pole is chosen to be $\omega_0<|p|<\omega_\mathrm{c}$ to increase its damping without adding significant phase lag around $\omega_\mathrm{c}$. 

Second, although the non-minimum phase zero is partially compensated, leading to a small phase lag in the crossover region only, further positive phase is needed. For this purpose, a lead-element is added to the PI-controller that yields
\vspace{-1.2ex}
\begin{equation}
K_\mathrm{PIL}(s) = \frac{213\, (s+2) \left(s+\frac{2}{3}\right)}{s\,(s+6)}.
\end{equation}

Note that Padé-term $\tilde{G}_\mathrm{d,1}$  has a stable pole that could be canceled. However, this would not lead to similar results, as it significantly reduces the magnitude slope and thus the damping of the ei\-gen\-frequency without adding enough positive phase in the crossover region.

For implementing the controller ${\tilde{C}_\mathrm{FO} = K_\mathrm{PIL}\, F_\mathrm{pp}\, \tilde Q^{-1}_{2/\tau_\mathrm{n},2}}$ and to use it in a real-time experimental setup, the FO-element $\tilde Q_{2/\tau_\mathrm{n},2}$ is approximated with an Oustaloup-Filter in the relevant frequency band $\left[\omega_\mathrm{l},\, \omega_\mathrm{h} \right] = \left[ 0.05\,\mathrm{\frac{rad}{s}},\, 50\,\mathrm{\frac{rad}{s}}\right] $ with $N = 1$, as described in \citep{Voss2022}. 

This approximation of the FO element leads to the overall controller $C_\mathrm{FO}$ of~\eqref{eq:controllerdesign-Cfo}.
A Bode plot of $L_\mathrm{FO} = C_\mathrm{FO} G$ is part of Fig.~\ref{fig:controllerdesign-bode}, where the open-loop crossover frequency is ${\omega_{\mathrm{c,FO}} = 1.49\,\mathrm{\frac{rad}{s}}}$, the phase and gain margins are ${\Phi_{\mathrm{r,FO}}= 37.36^\circ}$ and ${A_{\mathrm{r,FO}}= 1.81}$, respectively.

As the reference trajectory $ r(t)$ is known a priori, it can be differentiated offline and the classical feedforward 
\begin{equation}\label{eq:controllerdesign-Cfo-prefilter}
	u^\ast(t) = \frac{-CA^3 x^*(t) + r^{(3)}(t-\tau_\mathrm{n})}{CA^{2}B}
\end{equation}
can be applied, where the matrices $A$, $B$ and $C$ correspond to the model equations in~\eqref{eq:systemdynamics-ssr} and $x^*(t)$ to the state fitting the desired trajectory~\citep{Isidori2013}.

The open-loop feedforward path $F_\mathrm{FO}$ from~$r$ to~$u$ can be calculated by utilizing the Laplace transform on \eqref{eq:controllerdesign-Cfo-prefilter} for zero initial conditions, resulting in the non-proper
\begin{equation}
	F_\mathrm{FO}(s) = C_\mathrm{FO}(s) + G_\mathrm{yu}^{-1}(s)~e^{-\tau_\mathrm{n}s}.
\end{equation}
As the reference signal is known a-priori, this does not pose feasibility problems. Figure~\ref{fig:controllerdesign-bode-prefilter} holds a magnitude plot. It shows the dominance of $C_\mathrm{FO}$ for low frequencies and of the inverse plant for high frequencies, including the characteristic notch at $\omega_0$. In comparison with $C_\infty$, it is expected to act faster and more aggressive.

The controllers of the same order $n_\infty = n_\mathrm{FO}$ show very similar open-loop frequency characteristics for $r=0$ with $\omega_{\mathrm{c},\infty}\approx\omega_\mathrm{c,FO}$ but differences in the feedforward path, compare Figs.~\ref{fig:controllerdesign-bode} and~\ref{fig:controllerdesign-bode-prefilter}, respectively. 

\section{Simulation and Experimental Results}
\label{sec:results}
Two scenarios have been designed in order to evaluate the disturbance attenuation and the tracking performance. As the robustness against time delay uncertainties is of special interest, the communication delay is varied by 10~\%, i.e. $\tau = \tau_\mathrm{n}(1\pm0.1)$. The simulation model consists of the plant~\eqref{eq:systemdynamics-ssr}, the state-dependent input gain~$k_u\left(x_1\right)$ of \eqref{eq:systemdynamics-ku} and the communication delay~$\tau$. The experimental setup is depicted in Fig.~\ref{fig:schematics-experiment}.

Closed-loop experiments have revealed a necessary adjustment of the model input gain by $k_m =1.3$. This is addressed by the pre-compensation $k_m^{-1}$. Furthermore, a sticking phenomenon of the actuator similar to static friction is observed. Like in other fields, this can be addressed by injecting a high-frequency, low-amplitude jitter signal on top of the control signal, cf.~\citep{Haegglund2002} and~\citep{Hofer2018}. Best results have been achieved with a square wave with amplitude $A_\mathrm{j} = 0.4~\mathrm{V}$ and frequency $\omega_\mathrm{j} = 450~\mathrm{\frac{rad}{s}}$.

\subsection{Disturbance Attenuation}
\label{ssec:results-disturbance}
\begin{figure}[ht] 
	\centering
	\includegraphics[width=\linewidth]{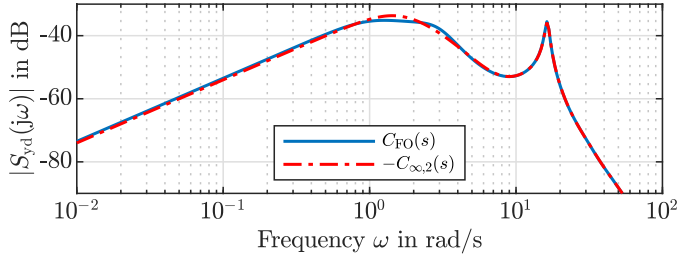}
	\caption{Magnitude plot of disturbance sensitivity functions.}
	\label{fig:results-sensitivity}
\end{figure}

In order to investigate the robustness against input disturbances, the disturbance sensitivity function
\begin{equation}
	S_\mathrm{yd}(s) = \frac{G_\mathrm{yd}(s)}{1+C(s)G_\mathrm{d}(s)G_\mathrm{yd}(s)}
\end{equation}
is depicted in Fig.~\ref{fig:results-sensitivity} for the nominal time delay of $\tau=\tau_\mathrm{n}$, where $C_\mathrm{FO}$ and $-C_{\infty,2}$ correspond to the FO loop-shaping and the 2DOF \hinf control, respectively. There are only slight differences in the crossover region leading to 
\begin{equation}
	\big\|S_\mathrm{yd,FO} \big\|_\infty = 0.0173
	\quad\text{and}\quad \big\| |S_{\mathrm{yd,}\infty} \big\|_\infty = 0.0203.
\end{equation}

The largest effect of the disturbance on the output is expected for frequencies $\omega\in \left[ 0.1\,\mathrm{\frac{rad}{s}},\, 20\,\mathrm{\frac{rad}{s}}\right]$. Therefore, a band- and amplitude-limited disturbance signal $d(t)$ is constructed from band-limited white noise via inverse Fourier transformation and normalization, such that 
the actuator input and states are prevented from saturation. 

\begin{figure}[ht] 
	\centering
	\includegraphics[width=1\linewidth]{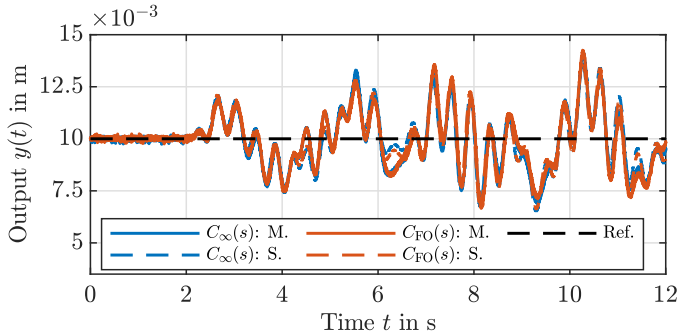}
	\caption{Exemplary input disturbance responses where `M.' abbreviates measurement and `S.'  simulation.}
	\label{fig:results-disturbance}
\end{figure}

The initial conditions for the disturbance attenuation experiments are chosen as $x_{\mathrm{0,d}} =  \left[0.01~0~0.01~0\right]^\top$ to be in the middle of the actuator range. Exemplary time responses for both controllers and $\tau=\tau_\mathrm{n}$ are illustrated in Fig.~\ref{fig:results-disturbance} (for a short period of time), where the disturbance injection starts at $t=2~\mathrm{s}$. There are only very little differences between the controllers, as expected by the similarity of the disturbance sensitivity functions in Fig.~\ref{fig:results-sensitivity}. Furthermore, the simulation results reasonably match the measurements, apart from short-time deviations (e.g. at~$t\approx 6.3~\mathrm{s}$).

\setbox0\hbox{\tabular{@{}c}Dist. atten.\endtabular}
\setbox1\hbox{\tabular{@{}c}Ref. track.\endtabular}
\begin{table}[ht]
	\centering
	\caption{Quantitative evaluation.}
	\label{tab:results-disturbance}
\setlength{\tabcolsep}{4pt}
	\begin{tabular}{l | lll|ccc}
		\textbf{Exp.} & \textbf{Contr.}	& {\textbf{Metric} } & \textbf{Unit}	&	$\tau=0.9\tau_\mathrm{n}$	&	$\tau=\tau_\mathrm{n}$	
																																		&	$\tau=1.1\tau_\mathrm{n}$\\ \hline
		\multirow{6}{*}{\rotatebox{90}{\usebox0}}		
		& $C_\infty$					&	$e_\mathrm{RMS} $	 &  mm				& \colorbox{green}{$1.624$}&$ 1.702 $&\colorbox{orange}{$1.729$}\\
		&								&	$\left|e_\mathrm{max} \right|$ & mm	& \colorbox{green}{$7.603$}&$ 7.986  $&$ 8.511$\\
		&								&  $\left|u_\mathrm{max} \right|$ &  V	& \colorbox{green}{$4.942$}&$ 4.974  $&$ 5.080$\\ \cline{2-7}
		& $C_\mathrm{FO}$		& 	$e_\mathrm{RMS} $	&  mm				& $ 1.637 $&$ 1.632$&$ 1.692 $ \\
		&								&	$\left|e_\mathrm{max} \right|$&  mm	& \colorbox{orange}{$8.812$}&$  8. 781  $&$ 8.683$\\
		&								&  $\left|u_\mathrm{max} \right|$& 	  V& $ 5.031  $&$ 5.010  $&\colorbox{orange}{$ 5.114$}\\ \hline\noalign{\smallskip}\hline
		\multirow{6}{*}{\rotatebox{90}{\usebox1}}		
		&$C_\infty$					&	$e_\mathrm{RMS} $		 &  mm			& \colorbox{orange}{$1.220$} &$ 1.215 $&$1.214 $\\
		&								&	$\left|e_\mathrm{max} \right|$  &  mm	& $4.911  $&\colorbox{orange}{$4.934$}&$ 4.907$\\
		&								&  $\left|u_\mathrm{max} \right|$ 	&  V&	$ 5.030 $ &$  5.047 $ &\colorbox{orange}{$5.061$}\\ \cline{2-7}
		&$C_\mathrm{FO}$		&$e_\mathrm{RMS} $ 					 &  mm&\colorbox{green}{$0.104$}&$ 0.149$&$ 0.221$\\
		&								&	$\left|e_\mathrm{max} \right|$  & mm	&\colorbox{green}{$0.474$}&$ 0.784  $&$ 1.205 $\\
		&								&  $\left|u_\mathrm{max} \right|$ 	 &  V	& $ 4.947  $&\colorbox{green}{$4.789$}&$ 4.813$\\ \hline
	\end{tabular}
\end{table}

In order to analyze the experimental data,
the root mean square~(RMS) error 
\vspace{-1.5ex}
\begin{equation}\label{eq:results-erms}\textstyle
	e_\mathrm{RMS} = \left(\frac{1}{N}\sum_{n=1}^{N}e^2_n\right)^\frac{1}{2},
\end{equation}
as well as the maximum absolute error and control effort
\begin{equation}\label{eq:results-max}
	\left|e_\mathrm{max} \right| = \max_{1\leq n\leq N}\left|e_n \right| \quad\text{and}\quad\left|u_\mathrm{max} \right| = \max_{1\leq n\leq N}\left|u_n \right| 
\end{equation}
are calculated, where $e_n = e(nT_\mathrm{s})$, $u_n = u(nT_\mathrm{s})$ and the sample time is $T_\mathrm{s}= 2\cdot 10^{-4}\,\mathrm{s}$. Table~\ref{tab:results-disturbance} summarizes the results. Obviously, $C_\infty$ performs the best for the reduced plant time delay $\tau=0.9\,\tau_\mathrm{n}$.  Focusing on the absolute values $\left|e_\mathrm{max} \right|$ and $\left|u_\mathrm{max} \right|$, it outperforms $C_\mathrm{FO}$. However, it shows more variations for a change in $\tau$ than $C_\mathrm{FO}$, especially for the RMS error $e_\mathrm{RMS} $. 
Still, the overall differences are rather small between the two controllers.

\subsection{Tracking Performance}
\label{ssec:results-tracking}
To analyze the tracking behavior, a transition polynomial
\begin{equation*}
	r(t) = r_0 + (r_1-r_0)  \left( \sum_{k=4}^{7} \frac{840\,(-1)^{k-2}}{k\, (k-2)!\, (7-k)!}\left(\frac{t}{T_\mathrm{t}}\right)^k\right)
\end{equation*}
of order~7 with vanishing derivatives at the boundaries for up to order~3 is utilized, i.e. a set-point change from ${r_0 = 5\,\mathrm{mm}}$ to ${r_1 = 10\,\mathrm{mm}}$ in ${T_\mathrm{t} = 2\,\mathrm{s}}$.

\begin{figure}
	\centering
	\includegraphics[width=\linewidth]{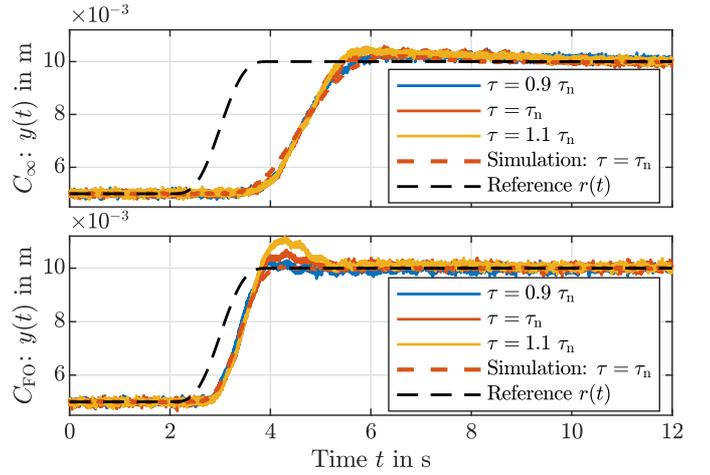}
	\caption{Time responses to the set-point change for both controllers (top: $C_\infty$, bottom: $C_\mathrm{FO}$).}
	\label{fig:results-stepresponse}
\end{figure}

Exemplary simulation and experimental results are shown in Fig.~\ref{fig:results-stepresponse}. The simulation results are given for the nominal time delay only (red dashed line) since the variation is very little. The set-point responses using 2DOF \hinf controller $C_\infty$~\eqref{eq:controllerdesign-Cinf},
turn out to be robust against time delay uncertainties. The results for the FO controller $C_\mathrm{FO}$~\eqref{eq:controllerdesign-Cfo} with prefilter~\eqref{eq:controllerdesign-Cfo-prefilter}
show a faster, however less robust behavior. This is mainly caused by the differences in the feed-forward paths, compare Fig.~\ref{fig:controllerdesign-bode-prefilter}.

\begin{figure}
	\centering
	\includegraphics[width=\linewidth]{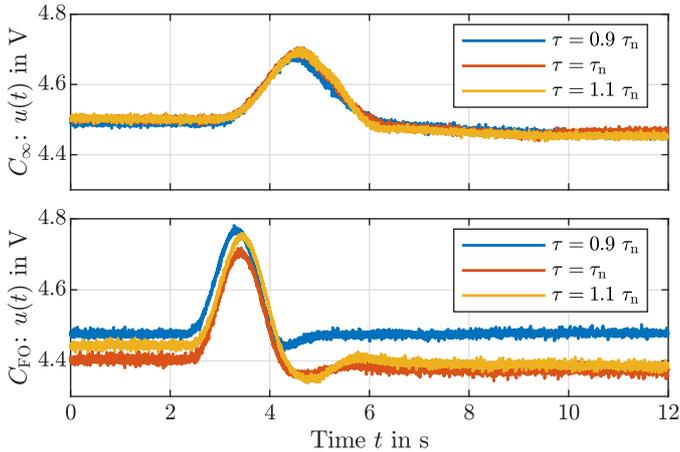}
	\caption{Control signals for the time responses in Fig.~\ref{fig:results-stepresponse}.}
	\label{fig:results-stepresponse-inputs}
\end{figure}
The control signals for the experiments are depicted in Fig.~\ref{fig:results-stepresponse-inputs} and confirm the expectations. The direct and more aggressive action of the FO controller is more affected by time delay uncertainties, whereas the behavior of the \hinf controller is slower but more robust. 

The quantitative evaluation of the results is summarized in Table~\ref{tab:results-disturbance} with the metrics defined in~\eqref{eq:results-erms} and~\eqref{eq:results-max}, averaged for three set-point changes. A time horizon of $20\,\mathrm{s}$ is used for the evaluation. Due to the different prefilters, there are significant differences between the two controllers for the error-related metrics, where $C_\mathrm{FO}$ with prefilter~\eqref{eq:controllerdesign-Cfo-prefilter} outperforms $C_\infty$.
However, $C_\infty$ is more robust against time delay uncertainties with respect to tracking performance leading to only slight variations of the metrics in contrast to $C_\mathrm{FO}$.  Here, the error related metrics increase significantly for an increasing time delay. The maximum control effort  $\left|u_\mathrm{max} \right|$, however, increases for any deviation from the nominal time delay due to the combination of the explicit time shift of the reference signal in $u^{\ast}(t)$ of \eqref{eq:controllerdesign-Cfo-prefilter} and the slower feedback control $C_\mathrm{FO}$.

\section{Conclusions}
\label{sec:conclusions}

The two designed controllers, a robust PI-Lead controller with FO elements and an IO \hinf controller, with similar frequency characteristics and controller orders show comparable results. Moreover, the simulations with a detailed model are in agreement with the experimental setup.

As expected from the magnitude plot of the input disturbance sensitivity function, the simulation and experimental results show very similar disturbance attenuation capabilities for both controllers. However, the \hinf controller turns out to be more robust against delay uncertainties. 

Significant differences are observed for the tracking performance due to the different feedforward actions. As expected from the magnitude plot, the FO controller with classical feedforward is more aggressive. Due to the explicit time-shift of the reference signal, it is sensitive to delay uncertainties. In contrast, the 2DOF \hinf controller is significantly slower, but generally more robust, as the delay uncertainties are not visible in the sample output graphs and the evaluation metric changes are marginal.

\section*{Acknowledgment}
This research was supported by IS-DAAD program under RCN project no. 320067
(DAAD project-ID 57458791). 

\bibliography{literature-FO-control}

\end{document}